\documentclass{article}
\usepackage{slashed,feynmf,epsfig,amsmath,amssymb,enumitem}
\usepackage[papersize={8.5in,11in}]{geometry}
\renewcommand{\vec}[1]{\boldsymbol{\mathrm{#1}}}
\usepackage{graphicx}
\newcommand{\be}{\begin{equation}}
\newcommand{\ee}{\end{equation}}
\begin{document}
\title{Modified Gravity (MOG), Cosmology and Black Holes}
\author{J. W. Moffat\\~\\
Perimeter Institute for Theoretical Physics, Waterloo, Ontario N2L 2Y5, Canada\\
and\\
Department of Physics and Astronomy, University of Waterloo, Waterloo,\\
Ontario N2L 3G1, Canada}
\maketitle




\begin{abstract}
A covariant modified gravity (MOG) is formulated by adding to general relativity two new degrees of freedom, a scalar field gravitational coupling strength $G= 1/\chi$ and a gravitational spin 1 vector field $\phi_\mu$. The $G$ is written as $G=G_N(1+\alpha)$ where $G_N$ is Newton's constant, and the gravitational source charge for the vector field is $Q_g=\sqrt{\alpha G_N}M$, where $M$ is the mass of a body. Cosmological solutions of the theory are derived in a homogeneous and isotropic cosmology. Black holes in MOG are stationary as the end product of gravitational collapse and are axisymmetric solutions with spherical topology. It is shown that the scalar field $\chi$ is constant everywhere for an isolated black hole with asymptotic flat boundary condition. A consequence of this is that the scalar field loses its monopole moment radiation.
\end{abstract}
\maketitle

\section{Introduction}

Dark matter was introduced to explain the stable dynamics of galaxies and galaxy clusters. General relativity (GR) with only ordinary baryon matter cannot explain the present accumulation of astrophysical and cosmological data without dark matter. However, dark matter has not been observed in laboratory experiments~\cite{Baudis}. Therefore, it is important to consider a modified gravitational theory. The difference between standard dark matter models and modified gravity is that dark matter models assume that GR is the correct theory of gravity and a dark matter particle such as WIMPS, axions and fuzzy dark matter are postulated to belong to the standard particle model. The present work introduces a simplified formulation of modified gravity (MOG), also called Scalar-Tensor-Vector-Gravity (STVG), that avoids unused generality of the original version~\cite{Moffat}. The MOG is described by a fully covariant action and field equations, extending GR by the addition of two gravitational degrees of freedom. The first is $G= 1/\chi$, where $G$ is the coupling strength of gravity and $\chi$ is a scalar field. The scalar field $\chi$ is motivated by the Brans-Dicke gravity theory~\cite{BransDicke,Weinberg,Narlikar}. The second degree of freedom is a massive gravitational vector field $\phi_\mu$. The gravitational coupling of the vector graviton to matter is universal with the gravitational charge $Q_g=\sqrt{\alpha G_N}M$, where $\alpha$ is a dimensionless scalar field, $G_N$ is Newton's gravitational constant and $M$ is the mass of a body.

We write $G=1/\chi$ as $G=G_N(1+\alpha)$ and $\chi$ is the only scalar field in the theory. The effective running mass of the spin 1 vector graviton is determined by the parameter $\mu$, which fits galaxy rotation curves and cluster dynamics without exotic dark matter~\cite{MoffatRahvar1,MoffatRahvar2,GreenMoffat,DavariRahvar}. It has the value $\mu\sim 0.01 - 0.04\,{\rm kpc}^{-1}$, corresponding to $\mu^{-1}\sim 25 - 100$ kpc and an effective mass $m_\phi\sim 10^{-26}-10^{-28}\,{\rm eV}$.
We derive generalized Friedmann equations and field equations for the scalar field $\chi$ in a homogeneous and isotropic universe with the Friedmann-Lema$\hat{i}$tre-Robertson-Walker (FLRW) line element and the energy-momentum tensor of a perfect fluid~\cite{Moffat2}.

We will consider black holes in MOG as the end point of gravitational collapse, which must be stationary axisymmetric and spherical solutions of the field equations. Following Hawking~\cite {Hawking}, the scalar field $\chi$ is constant everywhere in a stationary black hole solution and the black holes are described by the Schwarzschild-MOG and Kerr-MOG metric solutions~\cite{Moffat2,Moffat3}. An important consequence is that a MOG black hole will not have a scalar monopole moment. Moreover, due to the positivity of mass $M$ in the gravitational charge of the vector field, $Q_g=\sqrt{\alpha G_N}M$, a MOG black hole will not have a dipole moment.

In Section 2, we present the simplified MOG field equations and in Section 3, we review the equations of motion of a particle and the weak field approximation of the theory. In Section 4, we derive cosmological solutions and in Section 5, we investigate MOG black holes. We summarize the results in Section 6.

\section{The MOG Field Equations}

We will formulate the MOG action and field equations in a simpler and less general way than that first published in~\cite{Moffat}. We introduce $\chi= 1/G$ where $\chi$ is a scalar field and $G$ is the coupling strength of gravity~\cite{BransDicke,Weinberg,Narlikar}. The MOG action is given by (we use the metric signature (+,-,-,-) and units with $c=1$):
\be
S=S_G+S_\phi+S_M,
\ee
where
\be
S_G=\frac{1}{16\pi}\int d^4x\sqrt{-g}\biggl(\chi R+\frac{\omega_M}{\chi}\nabla^\mu\chi\nabla_\mu\chi +2\Lambda\biggr),
\ee
and
\be
\label{Baction}
S_\phi=\int d^4x\sqrt{-g}\biggl(-\frac{1}{4}B^{\mu\nu}B_{\mu\nu}+\frac{1}{2}\mu^2\phi^\mu\phi_\mu\biggr).
\ee
$S_M$ is the matter action and $J^\mu$ is the current matter source of the vector field $\phi_\mu$. Moreover, $\nabla_\mu$ denotes the covariant derivative with respect to the metric $g_{\mu\nu}$, $B_{\mu\nu}=\partial_\mu\phi_\nu-\partial_\nu\phi_\mu$ and $\omega_M$ is a constant. The Ricci tensor is
\be
R_{\mu\nu}=\partial_\lambda\Gamma^\lambda_{\mu\nu}-\partial_\nu\Gamma^\lambda_{\mu\lambda}
+\Gamma^\lambda_{\mu\nu}\Gamma^\sigma_{\lambda\sigma}-\Gamma^\sigma_{\mu\lambda}\Gamma^\lambda_{\nu\sigma}.
\ee
We expand $G$ by $G=G_N(1+\alpha)$, $\Lambda$ is the cosmological constant and $\mu$ is the effective running mass of the spin 1 graviton vector field.

Variation of the matter action $S_M$ yields
\be
T^M_{\mu\nu}=-\frac{2}{\sqrt{-g}}\frac{\delta S_M}{\delta g^{\mu\nu}},\quad J^\mu=-\frac{1}{\sqrt{-g}}\frac{\delta S_M}{\delta\phi_\mu}.
\ee
Varying the action with respect to $g_{\mu\nu}$, $\chi$ and $\phi_\mu$, we obtain the field equations:
\be
\label{Gequation}
G_{\mu\nu}=-\frac{\omega_M}{\chi^2}\biggl(\nabla_\mu\chi\nabla_\nu\chi -\frac{1}{2}g_{\mu\nu}\nabla^\alpha\chi\nabla_\alpha\chi\biggr)\\
-\frac{1}{\chi}(\nabla_\mu\chi\nabla_\nu\chi-g_{\mu\nu}\Box\chi)+\frac{8\pi}{\chi}T_{\mu\nu},
\ee
\be
\label{Bequation}
\nabla_\nu B^{\mu\nu}+\mu^2\phi^\mu=J^\mu,
\ee
\be
\label{Boxchi}
\Box\chi=\frac{8\pi}{(2\omega_M+3)}T,
\ee
where $G_{\mu\nu}=R_{\mu\nu}-\frac{1}{2}R$, $\Box=\nabla^\mu\nabla_\mu$, $J^\mu=\kappa\rho_M u^\mu$, $\kappa=\sqrt{G_N\alpha}$, $\rho_M$ is the density of matter and $u^\mu=dx^\mu/ds$. The energy-momentum tensor is
\be
T_{\mu\nu}=T^M_{\mu\nu}+T^\phi_{\mu\nu}+g_{\mu\nu}\frac{\chi\Lambda}{8\pi},
\ee
where
\be
T^\phi_{\mu\nu}=-\biggl({B_\mu}^\alpha B_{\alpha\nu}-\frac{1}{4}g_{\mu\nu}B^{\alpha\beta}B_{\alpha\beta}+\mu^2\phi_\mu\phi_\nu-\frac{1}{2}g_{\mu\nu}\phi^\alpha\phi_\alpha\biggr),
\ee
and $T=g^{\mu\nu}T_{\mu\nu}$. Eq.(\ref{Boxchi}) follows from the equation
\be
2\chi\Box\chi-\nabla_\mu\chi\nabla^\mu\chi=\frac{R}{\omega_M},
\ee
by substituting for $R$ from the contracted form of (\ref{Gequation}).

\section{Equations of Motion and Weak Field Approximation}

The equation of motion for a massive test particle in MOG has the
covariant form~\cite{Moffat,Roshan}:
\begin{equation}
\label{eqMotion}
m\biggl(\frac{du^\mu}{ds}+{\Gamma^\mu}_{\alpha\beta}u^\alpha
u^\beta\biggr)= q_g{B^\mu}_\nu u^\nu,
\end{equation}
where $u^\mu=dx^\mu/ds$ with $s$ the proper time along the particle
trajectory and ${\Gamma^\mu}_{\alpha\beta}$ denote the Christoffel
symbols. Moreover, $m$ and $q_g$ denote the test particle mass $m$ and
gravitational charge $q_g=\sqrt{\alpha G_N}m$, respectively.  For a
massless photon the gravitational charge vanishes,
$q_\gamma=\sqrt{\alpha G_N}m_\gamma=0$, so photons travel on null
geodesics $k^\nu\nabla_\nu k^\mu=0$~\cite{GreenMoffatToth}:
\begin{equation}
\frac{dk^\mu}{ds}+{\Gamma^\mu}_{\alpha\beta}k^\alpha k^\beta=0,
\end{equation}
where $k^\mu$ is the photon momentum and $k^2=k^\mu k_\mu=0$. We
note that for $q_g/m=\sqrt{\alpha G_N}$ the equation of motion for a
massive test particle (\ref{eqMotion}) {\it satisfies the (weak)
equivalence principle}, leading to the free fall of particles in a
homogeneous gravitational field, although the free-falling particles
do not follow geodesics.

In the weak field region, $r\gg 2GM$, surrounding a stationary mass
$M$ centred at $r=0$ the spherically symmetric field $\phi_\mu$, with
effective mass $\mu$, is well approximated by the Yukawa potential:
\begin{equation}
\phi_0=-Q_g\frac{\exp(-\mu r)}{r},
\end{equation}
where $Q_g=\sqrt{\alpha G_N}M$ is the gravitational charge of the source mass $M$.
The radial equation of motion of a non-relativistic test particle,
with mass $m$ and at radius $r$, in the field of $M$ is then given by
\begin{equation}
\label{Eq:MOGweakEOM}
\frac{d^2r}{dt^2}+\frac{GM}{r^2}=\frac{q_gQ_g}{m}\frac{\exp(-\mu r)}{r^2}(1+\mu r).
\end{equation}
The mass $\mu$ is tiny\,---\,comparable to the experimental bound on
the mass of the photon\,---\,giving a range $\mu^{-1}$ of the
repulsive exponential term the same order of magnitude as the size of
a galaxy.  Since $ q_gQ_g/m=\alpha G_NM$, the modified Newtonian
acceleration law for a point particle can be written
as~\cite{Moffat}:
\begin{equation}
\label{MOGaccelerationlaw}
a_{\rm MOG}(r)=-\frac{G_NM}{r^2}[1+\alpha-\alpha\exp(-\mu r)(1+\mu r)].
\end{equation}
This reduces to Newton's gravitational acceleration in the limit
$\mu r\ll 1$.

In the limit that $r\rightarrow\infty$, we get from
(\ref{MOGaccelerationlaw}) for approximately constant $\alpha$ and
$\mu$:
\begin{equation}
\label{AsymptoticMOG}
a_{\rm MOG}(r)\approx -\frac{G_N(1+\alpha)M}{r^2}.
\end{equation}
The MOG acceleration has a Newtonian-Kepler behaviour for large $r$ with enhanced
gravitational strength $G=G_N(1+\alpha)$. The transition from
Newtonian acceleration behavior for small $r$ to non-Newtonian
behaviour for intermediate values of $r$ is due to the repulsive
Yukawa contribution in (\ref{MOGaccelerationlaw}). This can also
result in the circular orbital rotation velocity $v_c$ having a
maximum value in the transition region.

For a distributed baryonic matter source, the MOG (weak field) acceleration law becomes:
\begin{equation}
\label{accelerationlaw2}
{a}_{\rm MOG}({\vec x})=-G_N\int d^3{\vec x}'
\frac{\rho_{\rm bar}({\vec x}')({\vec x}-{\vec x}')}{|{\vec x}-{\vec x}'|^3}
[1+\alpha-\alpha\exp(-\mu|{\vec x}-{\vec x}'|)(1+\mu|{\vec x}-{\vec x}'|)],
\end{equation}
where $\rho_{\rm bar}$ is the total baryon mass density.

A phenomenological formula for $\alpha$ for approximately constant $\alpha$ and weak gravitational field is~\cite{MoffatToth2009}:
\be
\label{alphaformula}
\alpha=\alpha_{\inf}\frac{M}{(\sqrt{M}+E)^2},
\ee
where $\alpha_{\inf}\sim{\cal O}(10)$ and $E=2.5\times 10^4\,M_{\odot}^{1/2}$. For $r\ll \mu^{-1}\sim 25 -100$ kpc the MOG acceleration reduces to the Newtonian acceleration $a_{\rm Newt}$. This is consistent with $\alpha\sim 10^{-9}$ obtained from (\ref{alphaformula}) for $M=1\, M_{\odot}$, guaranteeing that the solar system observational data is satisfied.

We have written the gravitational strength $G=G_N(1+\alpha)$, so we can relate the scalar field $\chi$ to this expansion of $G$ with $\alpha$ as $\chi\sim 1/G_N(1+\alpha)$. The constant $\omega_M$ can be set to $\omega_M=1$. In the weak gravitational acceleration formula (\ref{MOGaccelerationlaw}), the acceleration of a particle identified with a planet in the solar system approaches the Newtonian acceleration law as $r$ approached the size of the solar system. This is consistent with the phenomenological formula (\ref{alphaformula}) for $M\sim 1M_{\odot}$, yielding $\alpha\sim 10^{-9}$ and $\chi\sim 1/G_N$. In Brans-Dicke theory~\cite{BransDicke}, the constant $\omega$ is chosen to be large so that the theory can agree with solar system measurements. A prominent difference between Brans-Dicke gravity and MOG is the additional degree of freedom of the gravitational vector field $\phi_\mu$, allowing us to obtain the weak field MOG acceleration formula (\ref{MOGaccelerationlaw}). We have adopted a different approach for obtaining the solar system weak field limit by choosing the equivalent constant $\omega_M=1$ and the relation $\chi=1/G_N(1+\alpha)$. In the fitting of data when applying the field acceleration formula (\ref{MOGaccelerationlaw}) to {\it weak gravitational fields}, the parameters $\alpha$ and $\mu$ are treated as running constants that are not universal constants. The magnitude of $\alpha$ depends on the physical length scale or averaging scale $\ell$ of the system. For the solar system, $\ell_\odot\sim 0.5$ pc and for a galaxy $\ell_G\sim 5-24$ kpc.

\section{Cosmology}

We base our cosmology on the homogeneous and isotropic FLRW background metric:
\be
ds^2=dt^2-a^2(t)\biggl[\frac{dr^2}{1-Kr^2}+r^2(d\theta^2+\sin^2\theta d\phi^2)\biggr],
\ee
where $K=-1,0,1\,({\rm length}^{-2})$ for open, flat and closed universes, respectively. We use the energy-momentum tensor of a perfect fluid:
\be
\label{energymom}
T^{\mu\nu}=(\rho +p)u^\mu u^\nu-g^{\mu\nu}p.
\ee
We have ${T_0}^0=\rho$ and the density $\rho$ is
\be
\label{rho density}
\rho=\rho_M+\rho_r+\rho_\Lambda,
\ee
where $\rho_\Lambda=\Lambda/8\pi G_N$ and
\be
\rho_M=\rho_b+\rho_f.
\ee
Moreover, $\rho_f=\rho_\chi+\rho_\phi$ and $\rho_b,\rho_\chi,\rho_\phi$ denote the densities of baryon matter, scalar field matter and the electrically neutral gravitational vector field $\phi_\mu$, respectively. The radiation density $\rho_r=\rho_\gamma+\rho_\nu$, where $\rho_\gamma$ and $\rho_\nu$ denote the densities of photons and neutrinos, respectively.

In the following, we assume a spatially flat universe and we obtain the equations~\cite{Moffat2}:
\be
\label{Friedmann3}
\biggl(\frac{\dot a}{a}\biggr)^2=\frac{8\pi\rho}{3\chi}-\frac{{\dot\chi\dot a}}{\chi a}+\frac{\omega_M{\dot\chi}^2}{6\chi^2},
\ee
\be
\frac{\ddot a}{a}+\frac{{\dot a}^2}{2a^2}=-\frac{4\pi p}{\chi}
-\frac{\dot\chi\dot a}{\chi a}-\frac{\omega_M\dot\chi^2}{4{\chi}^2}-\frac{\ddot\chi}{2\chi},
\ee
\be
\label{chiequation}
\frac{1}{\chi^3}\frac{d}{dt}({\dot\chi a^3})=\frac{8\pi}{(2\omega_M+3)}(\rho-3p).
\ee
The energy conservation equation is
\be
\dot\rho+3\frac{d\ln a}{dt}(\rho+p)=0.
\ee

For the homogeneous and isotropic FLRW spacetime, only the $\phi_0$ component of $\phi_\mu$ is non-zero for $\mu\neq 0$. As for the photon radiation density $\rho_r$, we will treat the spin 1 graviton field particles as a gas described by the macroscopic fluid density $\rho_\phi$. The $\phi_\mu$ and $B_{\mu\nu}$ fields will enter the cosmological perterbation equations as fluctuates, $\delta\phi_\mu$ and $\delta B_{\mu\nu}$.

The integral of (\ref{chiequation}) is given by~\cite{Weinberg,Narlikar}:
\be
{\dot\chi}a^3=\frac{8\pi}{(2\omega_M+3)}\int dt(\rho-3p)\chi^3+C,
\ee
where $C$ is a constant. We obtain two types of solutions depending on whether $C=0$ or $C\neq 0$~\cite{Weinberg,Narlikar}.

Let us consider the case $C=0$. The matter dominated epoch has $p=0$ and we write
\be
\label{powerlaw}
a=a_0\biggl(\frac{t}{t_0}\biggr)^{n_1},\quad \chi=\chi_0\biggl(\frac{t}{t_0}\biggr)^{n_2}.
\ee
Then we have $\rho_M\propto t^{-3n_1}$ and the field equations give
\be
n_1=\frac{2\omega_M+2}{3\omega_M+4},\quad n_2=\frac{2}{3\omega_M+4},
\ee
and
\be
\rho_{M0}=\frac{(2\omega_M+3)n_2\chi_0}{8\pi t_0^2}.
\ee
We have for $\omega_M=1$, $n_1=4/7$, so there is a small deviation from the behavior in the matter dominated era, $a=a_0\biggl(\frac{t}{t_0}\biggr)^{2/3}$ and $\rho_M\propto 1/t^3$ of the GR Einstein-de Sitter universe model. Because $G= 1/\chi$, the time dependence of $G$ for $C=0$ is given by
\be
\frac{\dot G}{G}=-\frac{2}{3\omega_M+4}\frac{1}{t}=-\frac{H}{\omega_M+1},
\ee
where $H$ is the Hubble parameter.

For the case $C\neq 0$, the $\chi$ contributions dominate the dynamics of the early universe. We can choose values of $C$ that can describe both the matter dominated  era with $p=0$ and the radiation dominated era with $p=\frac{1}{3}\rho$. For a large enough $|C|$, the $\chi$ dominated solutions can differ significantly from the matter dominated solutions. Then, for $C$ large and negative $G$ can increase with time.

For the case of the flat Euclidean universe, $K=0$, $\rho_0=\rho_{\rm crit}$, and $\Omega_i=8\pi\rho_i/3\chi H^2$. We have
\be
\Omega_M=\frac{8\pi\rho_M}{3\chi H^2},\quad \Omega_\Lambda=\frac{8\pi\rho_\Lambda}{3\chi H^2},\quad \Omega_r=\frac{8\pi\rho_r}{3\chi H^2}.
\ee
We have two possible models for the universe. In the first model, it is assumed that $\rho_b < \rho_f=\rho_\chi+\rho_\phi$ in the early universe, followed by a transition to $\rho_b > \rho_f$ when the reionization period in the expansion of the universe begins and stars and galaxies are formed~\cite{Moffat2}. For this model, we can find values of $C\neq 0$ for the early universe and have $\rho_f > \rho_b$. In the late universe, following this period, MOG explains galaxy and galaxy cluster dynamics without dominant dark matter. In the second model, we assume that baryons dominate in the early and late time universe, $\rho_b > \rho_f$. For this model, we have $C=0$ and the baryons dominate the scalar and vector field contributions in both the early and late universe.

The first model can describe the cosmic microwave (CMB) data. The baryon sound wave oscillations due to the baryon-photon pressure prior to the decoupling time produce acoustical peaks in the angular power spectrum, ${\cal D}_\ell=\ell(\ell+1)C_\ell/2\pi$. We can match the $\Lambda$CDM calculation of the CMB angular acoustical power spectrum. The calculation of the power spectrum in the ${\Lambda}CDM$ model is duplicated in MOG, using the Planck 2018 Collaboration best-fit parameter values~\cite{Planck}: $\Omega_bh^2=0.0224\pm 0.0001$, $\Omega_fh^2=0.120\pm 0.001$, $\Omega_\Lambda=0.680\pm 0.013$, $n_s=0.965\pm 0.004$, $\sigma_8=0.811\pm 0.006$, $H_0=67.4\pm 0.5\,{\rm km}\,{\rm sec}^{-1}\,{\rm Mpc}^{-1}$, together with the remaining parameters in the fitting process. The second model has to fit the CMB data as well as the galaxy and cluster dynamics without exotic dark matter~\cite{MoffatToth2013}.

\section{MOG Black Holes}

For the matter-free $\phi_\mu$ field-vacuum case with $\Lambda=0$, $T^M_{\mu\nu}=0$ and $J^\mu=0$, the field equations are given by
\be
G_{\mu\nu}=-\frac{\omega_M}{\chi^2}\biggl(\nabla_\mu\chi\nabla_\nu\chi -\frac{1}{2}g_{\mu\nu}\nabla^\alpha\chi\nabla_\alpha\chi\biggr)\\
-\frac{1}{\chi}(\nabla_\mu\chi\nabla_\nu\chi-g_{\mu\nu}\Box\chi)+\frac{8\pi}{\chi}T^\phi_{\mu\nu},
\ee
\be
\label{Boxphivac}
\Box\chi=\frac{8\pi}{(2\omega_M+3)}T^\phi,
\ee
\be
\nabla_\nu B^{\mu\nu}+\mu^2\phi^\mu=0,
\ee
where
\be
T^\phi\equiv g^{\mu\nu}T^\phi_{\mu\nu}=\mu^2\phi^\mu\phi_\mu.
\ee

We will demonstrate in the following that for the black hole solution the vector field effective mass $\mu$ can be set to zero. A stationary black hole must be axisymmetric or static. In the former case there will be two Killing vectors fields, $\xi^\mu,\zeta^\mu$, the first is timelike and the second spacelike at infinity. A bivector $\xi^{[\alpha}\zeta^{\beta]}$ will be timelike at infinity and will have the magnitude
$f=\xi^{[\alpha}\zeta^{\beta]}\xi_{[\alpha}\zeta_{\beta]}$. An event horizon occurs when $f=0$. Outside the horizon at each point there exists a linear combination of $\xi^\mu$ and $\zeta^\mu$ which is timelike. According to Hawking~\cite{Hawking} the scalar field $\chi$ must be constant along the directions of $\xi^\mu, \zeta^\mu$, for they are Killing vectors. Therefore, in the exterior region the gradient of $\chi$ must be spacelike or zero everywhere. The same will be true in the static case, because there will be one Killing vector $\xi^\mu$ which is timelike everywhere in the exterior region. Let there be a partial Cauchy surface ${\cal S}$ for ${\bar J}^+({\cal J}^-)\cup{\bar J}^-({\cal J}^+)$ and a partial Cauchy surface ${\cal S}'$ determined by moving each point of ${\cal S}$ a unit distance along the integral curves of $\xi^\mu$. Let ${\cal V}$ be the region bounded by ${\cal S},{\cal S}'$, containing a portion of the event horizon and a timelike surface at infinity. The region exterior to the event horizon is empty apart from the energy-momentum tensor for the $\phi_\mu$ field.

We adopt the boundary condition $\chi=\chi_0$ where $\chi_0$ is the constant value at infinity. We now multiply (\ref{Boxphivac}) by $T^\phi$ and integrate over ${\cal V}$. Integrating by parts, we derive a volume integral over ${\cal V}$~\cite{Hawking,Faraoni}:
\be
\label{gradientchi}
\int_{\cal V}d^4x\sqrt{-g}T^\phi\Box\chi=\frac{8\pi}{(2\omega_M+3)}\int_{\cal V} d^4x\sqrt{-g}(T^\phi)^2,
\ee
together with surface integrals. Because of the isometry group the surface integral over the Cauchy surface ${\cal S}'$  cancels out the surface integral over ${\cal S}$. The surface integral at infinity is zero, because asymptotic flatness requires that the integral vanishes over the region at infinity for $\chi\rightarrow\chi_0$. The surface integral over the horizon is zero, because the gradient of $\chi$ is orthogonal to the null vector tangent to the horizon, which is a linear combination of the Killing vectors $\xi^\mu$ and $\zeta^\mu$. We obtain the result:
\be
\frac{8\pi}{(2\omega_M+3)}\int_{\cal V} d^4x\sqrt{-g}(T^\phi)^2=\frac{8\pi}{(2\omega_M+3)}\int_{\partial{\cal V}}d^3x\sqrt{|h|}T^\phi
\nabla_\mu\chi n^\mu=0,
\ee
where $\partial{\cal V}$ denotes the boundary of ${\cal V}$, $n^\mu$ is the normal to the boundary, and $h$ is the determinant of the induces metric $h_{\mu\nu}$ on the boundary. It follows that $T^\phi$ is manifestly zero. We have $T^\phi=\mu^2\phi^\mu\phi_\mu$, so for MOG black holes and $\phi_\mu\neq 0$ the effective mass $\mu=0$.

The conformally invariant gravitational $\phi$-field energy momentum tensor is
\be
\label{Bfieldenergy}
T^\phi_{\mu\nu}=-\biggl({B_\mu}^\alpha B_{\alpha\nu}-\frac{1}{4}g_{\mu\nu}B^{\alpha\beta}B_{\alpha\beta}\biggr),
\ee
and we have $T^\phi=g^{\mu\nu}T^\phi_{\mu\nu}=0$ giving the equation:
\be
\label{Boxchi2}
\Box\chi=0.
\ee

We now multiply (\ref{Boxchi2}) by $\chi$ and integrate over the volume ${\cal V}$. Integrating by parts, we derive the volume integral~\cite{Hawking}:
\be
\label{gradientchi}
\int_{\cal V}d^4x\sqrt{-g}\nabla^\mu\chi\nabla_\mu\chi=\int_{\partial{\cal V}}d^3x\sqrt{|h|}\nabla_\mu\chi n^\mu=0.
\ee
Because the volume integral (\ref{gradientchi}) must be zero and the gradient of $\chi$ is either spacelike or zero, then the gradient of $\chi$ must be zero everywhere and $\chi$ must be constant. In the original papers on MOG black holes~\cite{Moffat2,Moffat3}, we assumed that the gradient of $G$ is zero, so that $G=G_N(1+\alpha)=1/\chi$ is constant. We can now demonstrate that for MOG black holes, it is justified to have $G$ constant.

The metric for a static spherically symmetric black hole is given by~\cite{Moffat2,Moffat3}:
\be
\label{MOGmetric}
ds^2=\biggl(1-\frac{2G_N(1+\alpha)M}{r}+\frac{\alpha(1+\alpha)G_N^2M^2}{r^2}\biggr)dt^2-\biggl(1-\frac{2G_N(1+\alpha)M}{r}+
\frac{\alpha(1+\alpha)G_N^2M^2}{r^2}\biggr)^{-1}dr^2-r^2d\Omega^2,
\ee
where $d\Omega^2=d\theta^2+\sin^2\theta d\phi^2$. The metric reduces to the Schwarzschild solution when $\alpha=0$. The axisymmetric rotating black hole has the metric solution including the spin angular momentum $J=Ma$:
\be
ds^2=\frac{\Delta}{\rho^2}(dt-a\sin^2\theta d\phi)^2-\frac{\sin^2\theta}{\rho^2}[(r^2+a^2)d\phi-adt]^2-\frac{\rho^2}{\Delta}dr^2-\rho^2d\theta^2,
\ee
where
\be
\Delta=r^2-2G_N(1+\alpha)Mr+a^2+\alpha(1+\alpha)G_N^2M^2,\quad \rho^2=r^2+a^2\cos^2\theta.
\ee

Consider the timelike Killing vector $\xi^\mu$ in a stationary asymptotically flat solution of an uncollapsed body. The gravitational mass of the body measured at infinity is given by the $1/r$ and the $1/r^2$ contributions to the MOG metric in $\xi^\mu\xi^\nu g_{\mu\nu}$. We have
\be
M=\frac{1}{4}\pi\int d\Sigma_{\nu\beta}\nabla_\mu xg^{\mu\nu}\xi^\beta,
\ee
where $\Sigma_{\nu\beta}$ is the surface element of a spacelike 2-surface at infinity and $x^2=\xi^\mu\xi^\nu g_{\mu\nu}$. In the Einstein frame, $M$ denotes the gravitational mass calculated in the Einstein frame. Because the source mass $M$ in the gravitational charge of the vector field, $Q_g=\sqrt{\alpha G_N}M$, is positive the vector (spin 1 graviton) field does not produce a dipole moment. A consequence is that for an isolated black hole there will not be any dipole vector gravitational source, and because $\xi=1/G$ is constant there will not be any scalar monopole gravitational source. It can be demonstrated that for merging black holes $M_E$ will decrease by the amount of tensor (quadrupole) field energy radiated at infinity.

\section{Conclusions}

A simplified fully covariant modified gravity $MOG$ is formulated in terms of a scalar field $\chi= 1/G$ and a gravitational massive vector field $\phi_\mu$. The equations of motion of a particle satisfy the weak equivalence principle, because particles are in free fall but do not follow geodesics. In the weak field approximation, the MOG acceleration of a slowly moving particle has a repulsive Yukawa-type gravitational force in addition to the Newtonian gravitational force.

Cosmological solutions are derived for a homogeneous and isotropic universe. Two cosmological models are considered. The first model assumes that the density, $\rho_f=\rho_\chi+\rho_\phi$, mimics the role of dark matter in the early universe with $\rho_b < \rho_f$. A calculation of the CMB acoustical TT power spectrum is duplicated by MOG, using the basic six parameters in the $\Lambda CDM$ model~\cite{Planck}. A transition to a baryon dominated era, $\rho_b > \rho_f$, occurs at the time of reionization and the formation of the first stars and galaxies and MOG describes galaxy and galaxy cluster dynamics without a dominant dark matter. In the second model, it is assumed that baryons dominate in both the early and late universe, $\rho_b > \rho_f$~\cite{MoffatToth2013}.

The black holes as the end point of the gravitational collapse of an astrophysical body are stationary axisymmetric or static solutions of the matter free $\phi_\mu$-field vacuum MOG field equations with an asymptotic flat boundary condition. It follows that the scalar field $\chi$ must be a constant everywhere in a stationary MOG solution, and the gravitational coupling strength $G=G_N(1+\alpha)$ is constant. This has the consequence that as a body collapses, it loses its scalar monopole moment. The vector field $\phi_\mu$ dipole moment is also absent due to the positivity of the mass $M$ in the vector field gravitational source charge, $Q_g=\sqrt{\alpha G_N}M$. A mass loss results from the emission of tensor (quadrupole) radiation associated to the gravitational energy.

\section*{Acknowledgments}

I thank Martin Green and Viktor Toth for helpful discussions. Research at the Perimeter Institute for Theoretical Physics is supported by the Government of Canada through industry Canada and by the Province of Ontario through the Ministry of Research and Innovation (MRI).

\end{document}